%% file: main.tex
\PassOptionsToPackage{dvipsnames}{xcolor}
\documentclass[sigconf]{acmart}
\pdfoutput=1
\copyrightyear{2024}
\acmYear{2024}
\setcopyright{licensedusgovmixed}\acmConference[DAC '24]{61st ACM/IEEE Design Automation Conference}{June 23--27, 2024}{San Francisco, CA, USA}
\acmBooktitle{61st ACM/IEEE Design Automation Conference (DAC '24), June 23--27, 2024, San Francisco, CA, USA}
\acmDOI{10.1145/3649329.3657376}
\acmISBN{979-8-4007-0601-1/24/06}

\usepackage{amsmath}
\usepackage{mathtools}
\usepackage{graphicx}
\usepackage{textcomp}
\usepackage{xspace}
\usepackage{threeparttable}
\usepackage{siunitx}
\usepackage[inline]{enumitem}
\usepackage{subfig}

\usepackage{multirow}

\fancypagestyle{firstpage}{%
\lhead{}
\rhead{}
\fancyhead[CO]{Accepted for publication at the 61st Design Automation Conference (DAC '24). doi: \url{https://doi.org/10.1145/3649329.3657376}}
}

\begin{document}

\title[Synthesis of Resource-Efficient Superconducting Circuits with Clock-Free Alternating Logic]{Synthesis of Resource-Efficient Superconducting \\Circuits with Clock-Free Alternating Logic}

\author{Jennifer Volk}
\authornote{Both authors contributed equally to this research.}
\affiliation{
    \institution{University of Michigan}
    \city{Ann Arbor}
    \state{Michigan}
    \country{USA}
}
\email{jevolk@umich.edu}

\author{Panagiotis Papanikolaou}
\authornotemark[1]
\affiliation{
    \institution{University of Michigan}
    \city{Ann Arbor}
    \state{Michigan}
    \country{USA}
}
\email{panagip@umich.edu}

\author{Georgios Zervakis}
\affiliation{
    \institution{University of Patras}
    \city{Patras}
    \country{Greece}
}
\email{zervakis@upatras.gr}

\author{Georgios Tzimpragos}
\affiliation{
    \institution{University of Michigan}
    \city{Ann Arbor}
    \state{Michigan}
    \country{USA}
}
\email{gtzimpra@umich.edu}

\renewcommand{\shortauthors}{J. Volk, P. Papanikolaou, G. Zervakis, G. Tzimpragos}

\begin{abstract}
Gate-level clocking, typical in traditional approaches to  Single Flux Quantum (SFQ) technology, makes the effective synthesis of superconducting circuits a significant engineering hurdle. This paper addresses this challenge by employing the recently introduced alternating SFQ (xSFQ) logic family. xSFQ leverages dual-rail alternating encoding to eliminate the clock dependency from the superconducting gate semantics. This obviates the need for ad hoc modifications to existing synthesis tools and avoids unnecessary circuit resource overheads, marking a significant advancement in superconducting circuit design automation. Our implementation results demonstrate an average reduction of over 80\% in the Josephson junction count for circuits from the ISCAS85, EPFL, and ISCAS89 benchmark suites.
\end{abstract}


\maketitle
 \thispagestyle{firstpage}

\section{Introduction} 
\label{sec:intro}
\input{sections/introduction.tex}

\section{Standard Cell Library} 
\label{sec:sfq_lib}
\input{sections/xSFQ_lib.tex}

\section{Logic Synthesis \& Optimization} 
\input{sections/synthesis.tex}

\vspace{-1ex}
\section{Evaluation} 
\label{sec:eval}
\input{sections/evaluation.tex}

\section{Conclusion} 
\input{sections/conclusion.tex}

\bibliographystyle{ACM-Reference-Format}
\input{main.bbl}

\end{document}

%% file: sections/introduction.tex
Superconducting integrated circuits (ICs) offer the potential for a tenfold increase in performance while consuming a fraction of the power of today's semiconductor ICs, even considering cooling costs~\cite{7368011}. Their desirable characteristics also position superconductor electronics as critical enablers for next-generation breakthroughs in ultra-sensitive sensing and integrated classical-quantum computing~\cite{xqsim}. However, the path to practicality is still riddled with hurdles that need to be addressed, especially in design automation. 

In the most prominent superconducting technologies, computation evolves through the controlled propagation of single flux quanta, which can be conceptualized as picosecond-wide millivolt-level pulses. The presence of an SFQ pulse during a given interval is understood as a logical 1 and its absence a logical 0~\cite{rsfq}. This convention imposes two requirements: first, all logic gates must agree on an interval for evaluation; second, they must all be able to remember whether or not a pulse has arrived during the interval. The second requirement fits nicely with the inherently stateful nature of superconductor cells, but the first requirement is more difficult to fulfill and typically met by clocking every gate.

A closer examination of this fully synchronous reality reveals the delegation of up to $70\%$ of the circuit's Josephson junctions (JJs) for delay path balancing and clock signal splitting~\cite{sfqmap}; reduced operating speeds due to major metastability concerns; and limited architectural freedom, with the number of pipeline stages being dictated by the number of logic gates on the critical path. As for design automation, previous related research has primarily focused on exploring synchronization strategies, such as a single global clock or a combination of slow and fast clocks~\cite{10021969}, and customizing tools to restrict logical depth~\cite{pbmap}.
 
\begin{figure}[t!]
    \centering
    \includegraphics[width=0.95\linewidth]{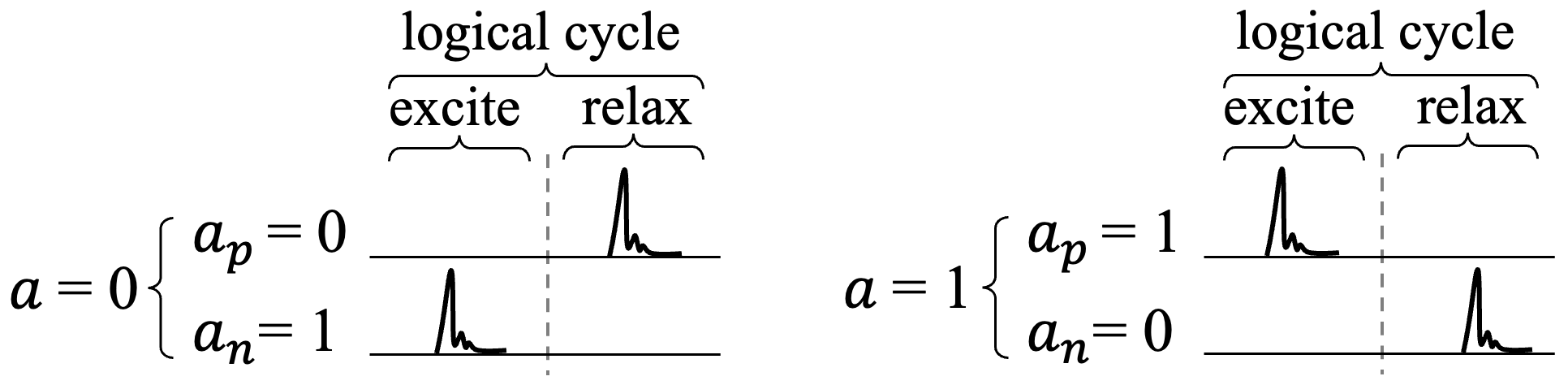}
    \vspace{-2mm}
    \caption{An xSFQ logical cycle consists of two synchronous phases (clock cycles): excite and relax. The value of an alternating binary variable appears during the excite phase and is followed by its complement during the relax phase.}
    \label{fig:enc}
\vspace{-2ex}
\end{figure}

This paper presents a radical departure from the above-described approaches. We showcase a synthesis framework for resource-efficient superconducting circuits from arbitrary register transfer level (RTL) code, employing exclusively clock-free logic gates and mature tools, without resorting to ad hoc modifications or the use of multiple clocks. To achieve this, we utilize the newly introduced alternating SFQ, xSFQ~\cite{xsfq}. Distinguishing itself from other superconductor logic families, xSFQ does not require an explicit clock signal~\cite{rsfq}, multi-phase AC biasing~\cite{pcl}, or flux leakage devices~\cite{dsfq} for evaluation and resetting. Rather, xSFQ repurposes easy-to-construct asynchronous cells---a Muller C element and its inverse---as AND and OR gates. We refer to each as the Last Arrival (LA) and First Arrival (FA) cells, borrowing terminology from the original paper. We accomplish completeness and cell reinitialization through \textit{dual-rail alternating encoding}---see Figure~\ref{fig:enc}.

Prior approaches to dual-rail codes in the context of superconductor electronics demonstrate data-driven self-timing~\cite{622205}. These efforts use complementary data inputs to generate timing signals and sidestep global clocking and distribution. However, this inflates cell size, from an average of 10 JJs in conventional SFQ approaches to 30, while still requiring delay path balancing. Additionally, a common misconception regarding dual-rail codes is the duplication of all hardware resources, as every dual-rail AND and OR gate now consists of a pair of AND and OR cells. 
This paper shows that, instead, dual-rail encoding can significantly enhance logic density by reducing the number of JJs per functional block.
To this end, in Section~\ref{sec:sfq_lib}, we 
\begin{enumerate*}[label=\bfseries(\alph*)]
\item present new cascadable LA and FA cell implementations, which require only 4 JJs each, and 
\item develop a complete xSFQ standard cell library for the synthesis of clock-free combinational and arbitrary sequential designs. 
Furthermore, we 
\item connect xSFQ to AND-Inverter graphs (AIG) and domino logic to exploit well-established optimizations and minimize the logic duplication penalty, as will be discussed in Section~\ref{sec:synthesis_A}.
\end{enumerate*}

Regarding the alternating aspect of xSFQ's encoding, every logical cycle consists of a pair of clock cycles, or synchronous phases. The excite phase operates on pulse-coded logic values and the relax phase operates on their pulse-coded complements. This alternation ensures that all LA/FA cells return to their initial state at the end of a logical cycle without the need for external signaling, as shown in Table~\ref{fig:alt_inp}. At the same time, though, this creates challenges. If a pair of synchronous destructive readout (DRO) cells is used as a storage element (e.g., flip-flop), the resulting circuit will have an unbalanced pipeline because no computation happens between the two of them~\cite{xsfq}. Moreover, if both DRO cells in a pair start with the same initialization, e.g., they both start at logical 0, this will violate the alternating property. In sequential designs, this can lead to malfunction as the logic cells will be out of phase and fail to properly reinitialize.
We address these two concerns by
\begin{enumerate*}[resume*]
\item leveraging traditional register retiming optimizations, such as those found in ABC, and
\item proposing an initialization strategy that guarantees correct excite and relax phase patterning.
\end{enumerate*}
These techniques are discussed in detail in Section~\ref{sec:synthesis_B}.

Lastly, for evaluation, we 
\begin{enumerate*}[resume*]
\item provide a combination of detailed analog and event-driven simulations and
\item present post-synthesis results for the ISCAS85, EPFL, and ISCAS89 circuit benchmark suites, covering both combinational and sequential designs.
\end{enumerate*}
A comparison with existing state-of-the-art techniques, presented in Section~\ref{sec:eval}, showcases up to a nearly 20$\times$ maximum reduction in the number of JJs and an average reduction of $4.3\times$, even when excluding cell interfacing, routing, and clock tree costs. As explained in the following sections, these ancillary costs are significantly lower in our xSFQ designs compared to conventional SFQ implementations.

\begin{table}[t!]
\caption{Alternating input sequences for the C element, or Last Arrival (LA), and its inverse, or First Arrival (FA), cells.}
\vspace{-3mm}
\footnotesize
\setlength{\tabcolsep}{3pt} 
\begin{tabular}{|c|c|c|c|c|c|c|c|c|c|c|}
\multicolumn{1}{c}{ } & \multicolumn{4}{c}{$\overbrace{\hphantom{ddddddddddddddd}}^\text{\normalsize excite}$} & \multicolumn{1}{c}{}& \multicolumn{4}{c}{$\overbrace{\hphantom{ddddddddddddddd}}^\text{\normalsize relax}$}  & \multicolumn{1}{c}{ } \\ 
\hline
\multirow{2}{*}{\textbf{state}} & \multicolumn{2}{c|}{\textbf{inputs}} & \multicolumn{2}{c|}{\textbf{outputs}} & \multirow{2}{*}{\textbf{state}} & \multicolumn{2}{c|}{\textbf{inputs}} & \multicolumn{2}{c|}{\textbf{outputs}} & \multirow{2}{*}{\textbf{state}} \\ \cline{2-5} \cline{7-10}
     &  \hspace*{0.2mm} a \hspace*{0.2mm} & \hspace*{0.2mm} b \hspace*{0.2mm} &  \textbf{FA}ab   &  \textbf{LA}ab   & &  \hspace*{0.2mm} a \hspace*{0.2mm} & \hspace*{0.2mm} b \hspace*{0.2mm} & \textbf{FA}ab  &  \textbf{LA}ab   & \\ \hline
Init &  0 & 0 & 0   & 0   & Init &  1 & 1 & 1   & 1   & Init \\ \hline
Init &  0 & 1 & 1   & 0   & b arrived &  1 & 0 & 0   & 1   & Init \\ \hline
Init &  1 & 0 & 1   & 0   & a arrived &  0 & 1 & 0   & 1   & Init \\ \hline
Init &  1 & 1 & 1   & 1   & Init &  0 & 0 & 0   & 0   & Init \\ \hline
\end{tabular}
\label{fig:alt_inp}
\end{table}

%% file: sections/xSFQ_lib.tex
\subsection{xSFQ logic gates}
\begin{figure} [t!]
    \centering
    \includegraphics[width=0.72\linewidth]{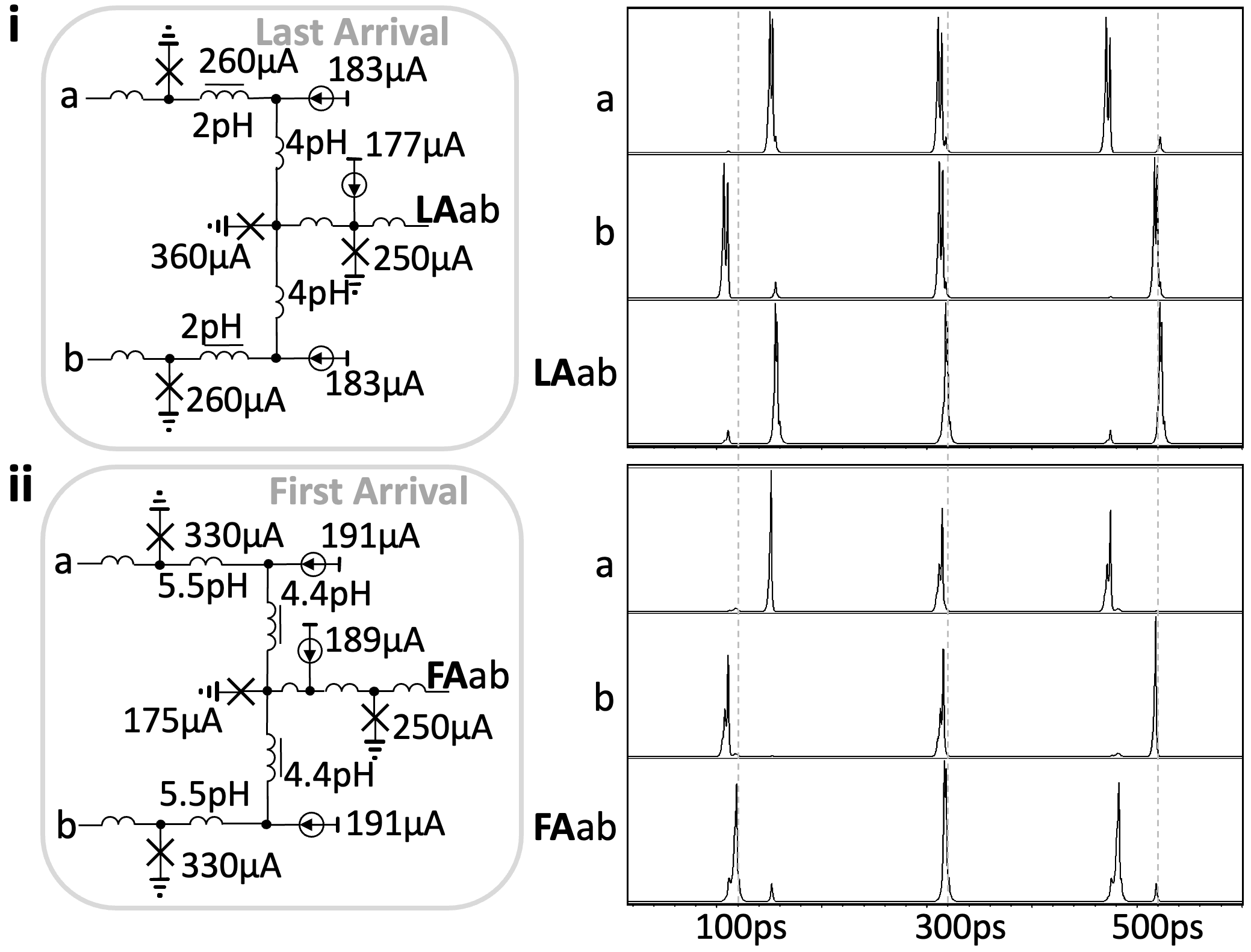}
    \vspace{-3mm}
    \caption{Schematics and SPICE waveforms for Last Arrival (Panel i) and First Arrival (Panel ii) cells.}
    \label{fig:FA-LA}\vspace{-3ex}
\end{figure}

The primary cells used for xSFQ logic operations are LA and FA. Existing implementations required 5 and 3 JJs for each, respectively~\cite{xsfq}. Here, we simplify the design of the LA cell by removing the dynamic resetting loops, which are not necessary. Additionally, we incorporate an extra JJ in the output of both LA and FA cells to enhance their cascadability by adhering to I$_C$ cell ranking rules~\cite{volk2023low}. Both cells' schematics and corresponding SPICE-level simulations are shown in Figure~\ref{fig:FA-LA}. The LA cell generates an output upon receiving pulses on both $a$ and $b$ input lines and transitions back to its original state. The FA cell produces an output upon receiving the first input pulse on either the $a$ or $b$ line and holds state until the second input pulse arrives and triggers a return to its initial state. 

\subsection{xSFQ storage elements}
In xSFQ, combinational logic no longer requires clocked gates. However, clocked storage elements (e.g., flip flops) are still required for sequential logic. In previous literature, this role was played by two pairs of DROs in order to respect xSFQ's dual-rail alternating encoding~\cite{xsfq}. Here, we present an alternative storage element, a single pair of DRO cells with complementary outputs (DROC), which, as will be described in Section~\ref{sec:synthesis_A}, creates opportunities for reducing the logic duplication penalty associated with dual-rail circuits.

\begin{figure} [ht]
    \centering
    \includegraphics[width=0.86\linewidth]{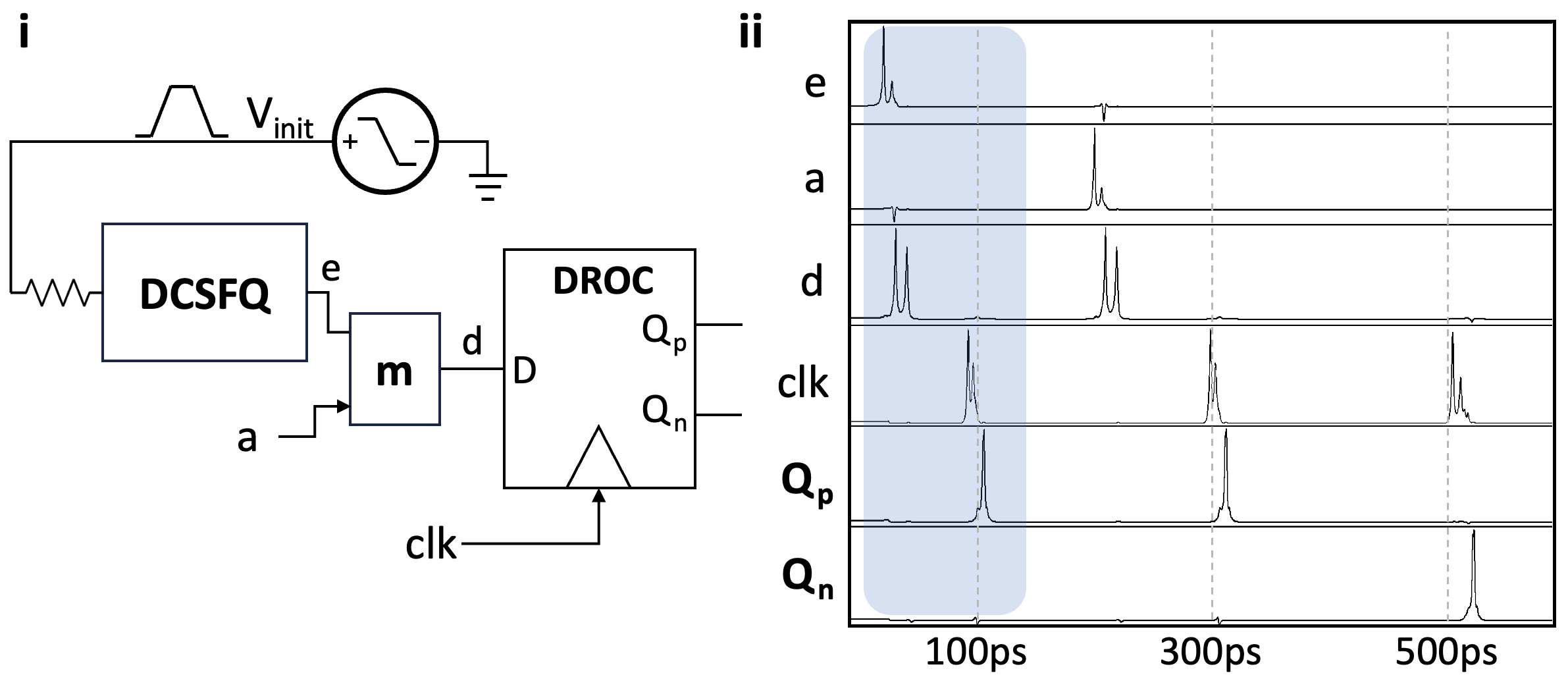}
    \vspace{-2mm}
    \caption{Block diagram (Panel i) and SPICE waveform (Panel ii) for a DROC cell with DC-to-SFQ preloading hardware. Symbol $m$ denotes a merger. The highlighted part of the waveform shows how we externally set the DROC cell and its response to the arriving clock pulse.}
    \label{fig:DROC}
    \vspace{-3ex}
\end{figure}

An important detail, overlooked so far by previous xSFQ literature, is that in the case of sequential logic, the two synchronous DRO/DROC cells that form a logical storage element must never both return 0 or 1 at the same time, as this violates the alternating requirement. We address this issue by designing a DROC cell that can output a logical 1 during its first operational cycle. A straightforward approach to preloading a DROC cell is to merge an external SFQ signal with the DROC's data input. However, distributing such a signal to all necessary DROC cells incurs resource overhead due to splitting and routing. To preload a DROC cell without these costs, we utilize a DC-to-SFQ converter composed of 4 JJs and connected to a global voltage line, which does not incur the same routing and splitting overheads as an SFQ-carrying line. A block diagram and corresponding SPICE-level simulation results are shown in Figure~\ref{fig:DROC}. For cases where the preloading is not needed, the extra hardware is removed; the synthesized designs use both such DROC cells.

\subsection{Library characterization}
For standard cell building and library characterization, we utilize the common JJ model from the MIT Lincoln Laboratory SFQ5ee 100 $\mu$A/$\mu$m$^2$ process~\cite{tolpygo2016advanced}. We use Synopsys' HSPICE simulator to extract propagation and clock-to-Q delays from JJ phase rise times. As stated in previous research~\cite{10021969}, the use of passive transmission lines (PTLs) for routing reduces the timing arc information to single values;
thus, after performing cell characterization, we format timing data into 1\raisebox{0.1 em}{\scalebox{0.7}{$\times$}}1 look-up tables. 

PTLs are passive interconnects typically used for cell connectivity at intermediate distances; they require designated drivers and receivers that increase the JJ count. For the comparative cases without PTLs, we average the propagation delays for each cell across all preceding and succeeding load combinations, reflecting the impact of load variations on timing~\cite{9360505}. Our Liberty file includes latency and JJ count information for all cells listed in Table~\ref{tab:cells}. Two DROCs are considered: one with hardware for preloading and one without.

\begin{table}[b]
\caption{Delays and JJ counts for xSFQ logic, storage, and fanout cells. DROC preloading hardware accounts for 9 JJs.}
\vspace{-2mm}
\label{tab:cells}
\footnotesize
\begin{tabular}{lcccc}
\multicolumn{1}{c}{ } & \multicolumn{2}{c}{$\overbrace{\hphantom{ddddddddddddddd}}^\text{\small without PTLs}$} & \multicolumn{2}{c}{$\overbrace{\hphantom{ddddddddddddddd}}^\text{\small with PTLs}$} \\ \hline
\multicolumn{1}{|l|}{\textbf{Cell}} & \multicolumn{1}{c|}{\textbf{Delay (ps)}} & \multicolumn{1}{c|}{\textbf{\# JJs}} & \multicolumn{1}{c|}{\textbf{Delay (ps)}} & \multicolumn{1}{c|}{\textbf{\# JJs}} \\ \hline
\multicolumn{1}{|l|}{JTL}           & \multicolumn{1}{c|}{4.6}                 & \multicolumn{1}{c|}{2}               & \multicolumn{1}{c|}{17}                  & \multicolumn{1}{c|}{7}               \\ \hline
\multicolumn{1}{|l|}{LA}     & \multicolumn{1}{c|}{7.2}                 & \multicolumn{1}{c|}{4}               & \multicolumn{1}{c|}{19.9}                & \multicolumn{1}{c|}{12}              \\ \hline
\multicolumn{1}{|l|}{FA} & \multicolumn{1}{c|}{9.5}                 & \multicolumn{1}{c|}{4}               & \multicolumn{1}{c|}{24.7}                & \multicolumn{1}{c|}{12}              \\ \hline
\multicolumn{1}{|l|}{DROC (Q$_p$)}  & \multicolumn{1}{c|}{6.7}                 & \multicolumn{1}{c|}{13/22}           & \multicolumn{1}{c|}{18}                  & \multicolumn{1}{c|}{27/36}              \\ \hline
\multicolumn{1}{|l|}{DROC (Q$_n$)}  & \multicolumn{1}{c|}{9.5}                 & \multicolumn{1}{c|}{13/22}           & \multicolumn{1}{c|}{21.5}                & \multicolumn{1}{c|}{27/36}              \\ \hline
\multicolumn{1}{|l|}{Splitter}         & \multicolumn{1}{c|}{5.1}                 & \multicolumn{1}{c|}{3}               & \multicolumn{1}{c|}{19.7}                & \multicolumn{1}{c|}{10}              \\ \hline
\end{tabular}
\end{table}

%% file: sections/synthesis.tex
\label{sec:synthesis}
\subsection{Circuits without synchronous elements}
\label{sec:synthesis_A}

\subsubsection{Direct RTL-to-xSFQ mapping}
\label{synth:default}
To translate arbitrary RTL code to an xSFQ gate netlist, we initially perform synthesis as in a typical CMOS setting and then replace every AND and OR gate in the resulting netlist with its dual-rail equivalent: a pair of LA-FA cells. For inversion, all that is needed is to ``twist'' the complementary wires. This process produces functionally correct circuits but approximately doubles the number of required cells. As an example, a full adder is traditionally implemented with 2 XOR gates, 2 AND gates, and 1 OR gate, or equivalently, with 9 NAND gates. When mapped to LA-FA pairs, the resulting xSFQ netlist comprises 18 LA/FA cells and 16 splitters for fanout, which translates to 120 JJs without PTL interfaces and 264 JJs with them\footnote{We assume that cell abutment can be performed at the outputs of the splitters, and thus we exclude the associated cost of PTL drivers and receivers.}. The use of clock-free cells eliminates synchronization overhead.

\subsubsection{Splitters minimization}
In this paper, we assume that all logic operators, LA and FA, have two inputs and one output. Additionally, we only consider splitter cells that support fanouts of two. Under these assumptions, the number of splitters is given by Equation~\ref{eq:splt}. 

\vspace{-2mm}
\small
\begin{equation}
\label{eq:splt}
\begin{split}
N_{splt} &=  \overbrace{(2N_{gate}+N_{out})}^{\text{\scriptsize connections needed}} - \overbrace{(N_{gate}+N_{inp})}^{\text{\scriptsize signals available}} \\
&= N_{gate}+N_{out}-N_{inp}.
\end{split}
\end{equation}\normalsize

\noindent$N_{out}$ and $N_{inp}$ represent the number of outputs and inputs of the combinational logic block, respectively. $N_{gate}$ denotes the number of LA and FA cells. Consequently, minimizing the number of LA and FA cells also implicitly minimizes the required splitters and the total JJ count in general. 

\subsubsection{LA-FA pairs minimization with AIG optimizations}
\begin{figure}[t]
    \centering
    \includegraphics[width=0.93\linewidth]{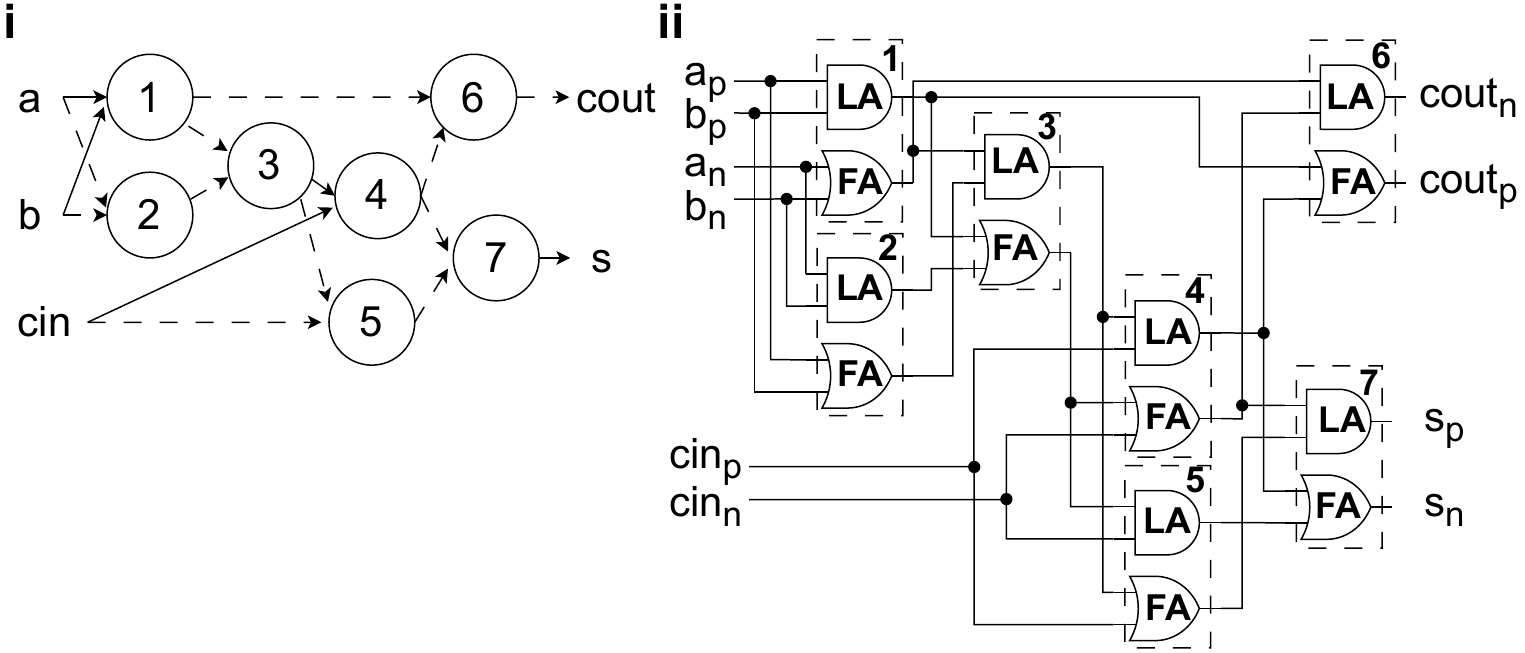}
    \vspace{-3mm}
    \caption{Minimal AIG of a full adder (Panel i). AIG nodes represent AND operations, while dotted lines indicate inversions. Isomorphic xSFQ circuit (Panel ii). Subscripts $p$ and $n$ denote the positive and negative polarities, respectively, of each dual-rail signal.}
    \label{fig:adder_aig}\vspace{-3ex}
\end{figure}

We notice that a dual-rail xSFQ circuit composed of LA-FA pairs is isomorphic to a single-rail CMOS circuit with AND and NOT gates. To establish this isomorphism, we replace the LA cell in each LA-FA pair with an AND gate. To derive the FA's output, we connect an inverter to the AND gate's output because the FA cell always generates the opposite signal of the LA cell\footnote{This process can be reversed to map an AND-Inverter graph into an xSFQ circuit.}. The resulting circuit resembles an AIG~\cite{aig}, which is a representation commonly used for logic optimization in tools such as ABC. Therefore, reducing the number of LA-FA pairs in an xSFQ netlist is equivalent to minimizing the number of nodes in the corresponding AIG. For instance, in Figure~\ref{fig:adder_aig}, the minimal AIG of a full adder consists of 7 nodes, as derived by ABC, translating to 14 LA/FA cells. This is a reduction of 2 AIG nodes, or 4 LA/FA cells and 4 splitters, and saves $28/60$ JJs (without and with PTL overheads) compared to the original design described in Section~\ref{synth:default}. 

Note that while ABC has been employed in the past for the synthesis of rapid SFQ (RSFQ) circuits, it required significant customization~\cite{pbmap}. In contrast, xSFQ netlists exhibit seamless compatibility with ABC's internal AIG representation. To the best of our knowledge, this is the first effort that achieves the synthesis of superconducting circuits with direct application of off-the-shelf AIG optimizations.

\subsubsection{LA/FA cells minimization with polarity optimizations}\label{sec:fixedpol}
Without compromising generality, we can assume that a primary output of a combinational xSFQ circuit either drives a synchronous element such as a DROC cell or connects to an alternating dual-rail to single-rail converter~\cite{xsfq}. Consequently, while xSFQ requires dual-rail encoding to fulfill completeness requirements, this constraint does not strictly apply to the circuit's outputs. 

We exploit the relaxation of dual-rail encoding at the outputs by retaining only the positive polarity. This entails starting from the negative polarity outputs and systematically eliminating redundant logic as we traverse the circuit in reverse. This technique resembles backward bubble pushing, in which inverters are propagated from outputs to inputs in an AIG by invoking De Morgan's rule when necessary. Once a parsed AIG node no longer has inverted outputs, it can be mapped to either a single LA or FA cell, depending on whether De Morgan's rule was applied to this node. Specifically, an AND node maps to an LA cell, and an equivalent OR node maps to an FA cell. LA-FA pairs are only required for nodes on which inverters cannot propagate and for the logic that precedes them. An inverter becomes impeded by a gate when the gate has a fanout greater than one and two or more outputs that are complementary. 

\begin{figure}[t!]
    \centering
    \includegraphics[width=\linewidth]{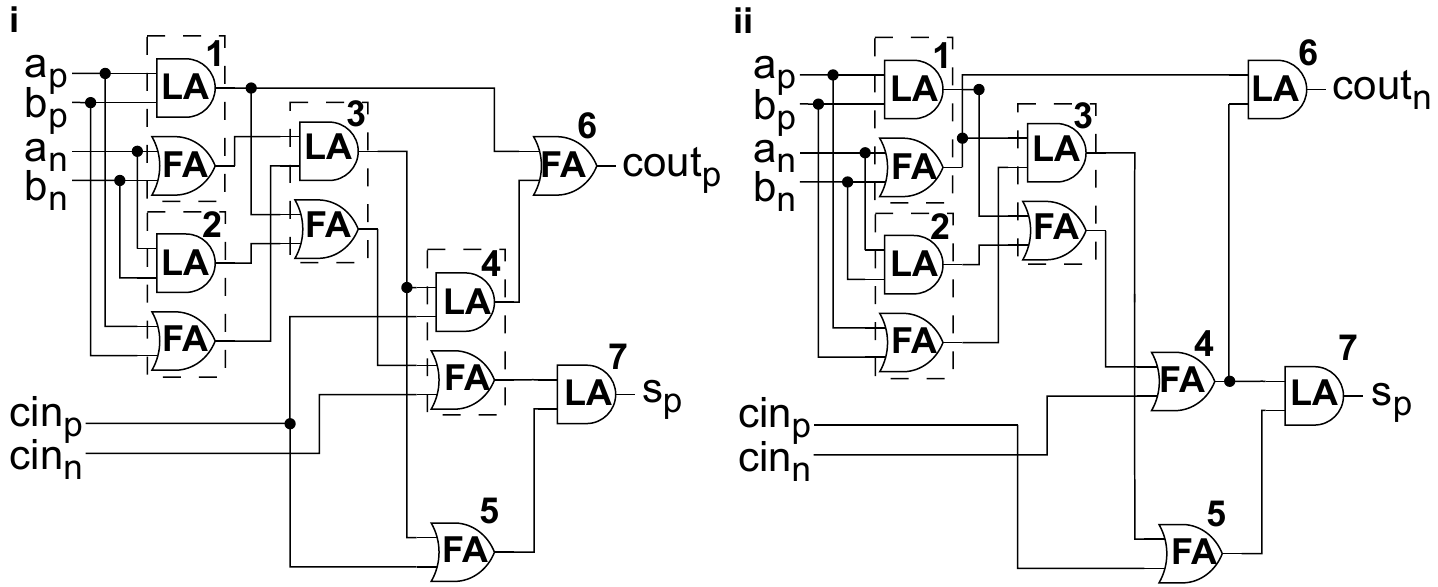}
    \vspace{-5mm}
    \caption{Full adder circuits consisting of 11 (Panel i) and 10 (Panel ii) LA/FA cells following our polarity optimizations.}
    \label{fig:adder_polarity}\vspace{-2ex}
\end{figure}

Applying this optimization to our example full adder circuit results in 11 LA/FA cells, as depicted in Figure~\ref{fig:adder_polarity}i. This results in a further reduction of 3 LA/FA cells and 5 splitters---a savings of $27/51$ JJs---compared to the AIG-optimized design. The reduction is achieved by eliminating the computation of $cout_n$ and $s_n$, which simplifies three out of the seven LA-FA pairs to single LA/FA cells. 

\subsubsection{xSFQ like domino logic.}\label{sec:domino}
\noindent Figure~\ref{fig:adder_polarity}ii illustrates an even more compact full adder implementation by retaining $cout_n$ instead of $cout_p$. This optimized design employs only 10 LA/FA cells and 6 splitters, or $58/138$ JJs---a $50\%$ reduction from the initial implementation, which comprises $120/264$ JJs. In general, selecting the negative instead of the positive polarity on a primary output does not impose any expressivity constraints due to the interfacing assumptions made above, such as the use of DROC cells at the boundaries of the combinational logic. To systematize the search for the combination of signal polarities that minimize LA/FA cell count, we establish a connection between xSFQ and domino logic. 

Akin to xSFQ, domino logic is inherently non-inverting. This essentially implies that xSFQ and domino logic encounter the same risk of doubling the number of required cells if intermediate inversions in the corresponding AIG are not minimized, as described in Section~\ref{sec:fixedpol}. Prior domino logic literature offers a heuristic solution to the output polarity assignment problem for minimum logic duplication~\cite{mininv, outopt}. We apply the same heuristic here to reduce the number of required LA/FA cells.

Results for the EPFL benchmark control circuits are presented in Table~\ref{tab:duplication}. The duplication penalty is generally well-constrained, with some cases as low as 0\%---compared to 100\% in the case of direct mapping discussed in Section~\ref{synth:default}. An exception is the voter circuit, which exhibits a high rate. To the best of our understanding, this is due to the given implementation, which features an output comparator that requires both polarities from its inputs. This provides few opportunities to apply the optimizations described above; however, we derived an alternative implementation in sum-of-products form that has a 0\% penalty. More detailed results and comparisons against the state of the art in terms of JJ count are provided in Section~\ref{sec:eval}.

\subsection{Circuits with synchronous elements} 
\label{sec:synthesis_B}

\begin{figure}[t]
    \centering
    \includegraphics[width=0.97\linewidth]{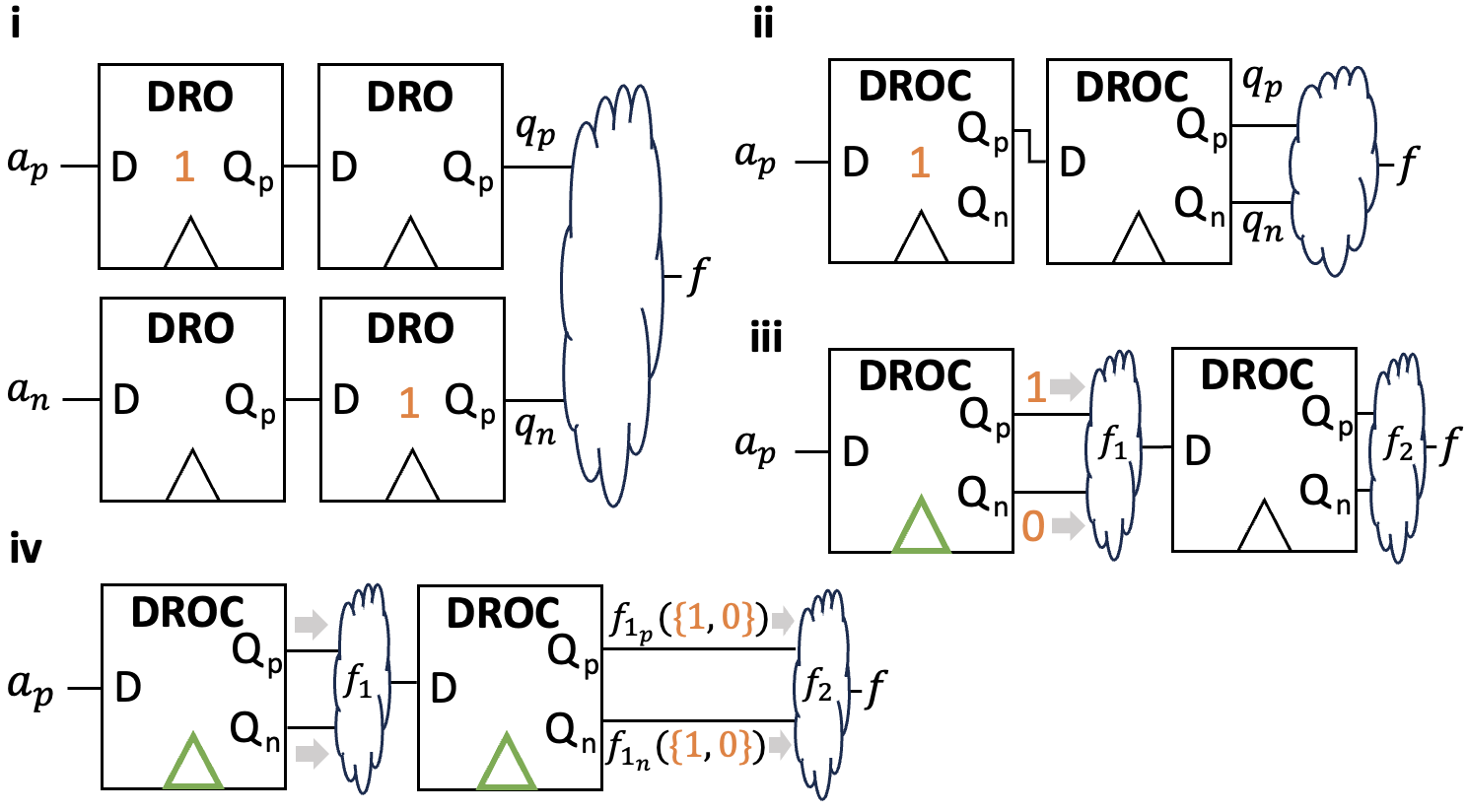}
    \vspace{-2mm}
    \caption{Original xSFQ logical flip-flop with two DRO cells per polarity (Panel i). Preloading is indicated by orange `1's. Proposed implementation with two DROC cells (Panel ii). Retimed design for pipeline balancing (Panel iii) with the preceding DROC cell clocked first (green clock port). The circuit is then ready for normal operation (Panel iv).}
    \label{fig:sync}\vspace{-3ex}
\end{figure}

Combinational xSFQ logic designs are, by nature, clock-free and do not necessitate the use of any synchronous components. However, in the case of sequential logic, synchronous storage elements are needed. According to literature~\cite{xsfq}, to adhere to xSFQ's dual-rail alternating encoding, every logical flip-flop should consist of four DRO cells, as shown in Figure~\ref{fig:sync}i. More precisely, two DRO cells in a pair satisfy the alternating property, and utilizing two such DRO pairs fulfills the dual-rail property.  

In the context of feedforward xSFQ circuits, preloading any of these four DRO cells is not necessary. If all DRO cells within the same synchronization stage return no pulse, and the successive LA/FA cells are in their initial state, no malfunction will occur. However, in circuits featuring feedback loops, this approach will result in a mix of alternating and non-alternating encoded data. Due to this mix, some LA/FA cells may fail to properly reinitialize by the end of a logical cycle---Table~\ref{fig:alt_inp}. Our solution is to preload two out of the four DRO cells, indicated in Figure~\ref{fig:sync}i by orange 1s.

\begin{table}[t!]
\caption{Duplication penalty for the EPFL benchmark control circuits after the proposed optimizations. }
\vspace{-2mm}
\footnotesize
\begin{tabular}{cccccc}
\hline
\multicolumn{1}{|c|}{\textbf{Circuit}} & \multicolumn{1}{c|}{arbiter}   & \multicolumn{1}{c|}{calvc}     & \multicolumn{1}{c|}{ctrl}     & \multicolumn{1}{c|}{dec}    & \multicolumn{1}{c|}{i2c}   \\ \hline
\multicolumn{1}{|c|}{\textbf{Dupl.}}   & \multicolumn{1}{c|}{0\%}       & \multicolumn{1}{c|}{8\%}       & \multicolumn{1}{c|}{9\%}      & \multicolumn{1}{c|}{0\%}    & \multicolumn{1}{c|}{6\%}   \\ \hline
\\[-2mm] \hline
\multicolumn{1}{|c|}{\textbf{Circuit}} & \multicolumn{1}{c|}{int2float} & \multicolumn{1}{c|}{mem\_cnrl} & \multicolumn{1}{c|}{priority} & \multicolumn{1}{c|}{router} & \multicolumn{1}{c|}{voter} \\ \hline
\multicolumn{1}{|c|}{\textbf{Dupl.}}   & \multicolumn{1}{c|}{6\%}       & \multicolumn{1}{c|}{6\%}       & \multicolumn{1}{c|}{22\%}     & \multicolumn{1}{c|}{44\%}   & \multicolumn{1}{c|}{99\%}  \\ \hline
\end{tabular}
\label{tab:duplication}
\vspace{-2ex}
\end{table}

Figure~\ref{fig:sync}ii introduces a new design for the xSFQ logical flip-flop that utilizes two DROC cells instead of four DRO cells. The transition from DRO to DROC cells leads to a smaller clock tree and enables the optimizations discussed in Sections~\ref{sec:fixedpol} and \ref{sec:domino}. Following the same rationale as above, the first DROC cell should be preloaded and the second should not. This setup works as expected for any sequential design, but from an architectural standpoint, there is a noticeable pipeline imbalance---no computation happens between the two successive synchronous cells. We address this concern by splitting DROC pairs and pushing individual DROC cells through the fabric of combinational xSFQ logic using ABC's retiming capabilities.

The same circuit with a rebalanced pipeline, $f$ split into $f_1$ and $f_2$, is depicted in Figure~\ref{fig:sync}iii. However, pushing combinational logic between the two DROC cells that previously formed a pair may pose a challenge. LA and FA cells have an internal state that persists between clock cycles. If all DROC cells generate an output pulse during the first clock cycle, either through their $Q_p$ or $Q_n$ ports, the succeeding LA/FA cells will exit their initial phase and enter their excite phase. This is a violation because, for correct pipelined operation, two consecutive combinational logic blocks separated by a synchronous barrier cannot both be in the same phase~\cite{xsfq}. Disregarding this violation will lead to erroneous operation for $f_2$, as the subsequent clock cycle will find its LA/FA cells in their a or b arrived states (Table~\ref{fig:alt_inp}) instead of their initial states.

\begin{figure} [b]
\vspace{-2ex}
    \centering    \includegraphics[width=0.95\linewidth]{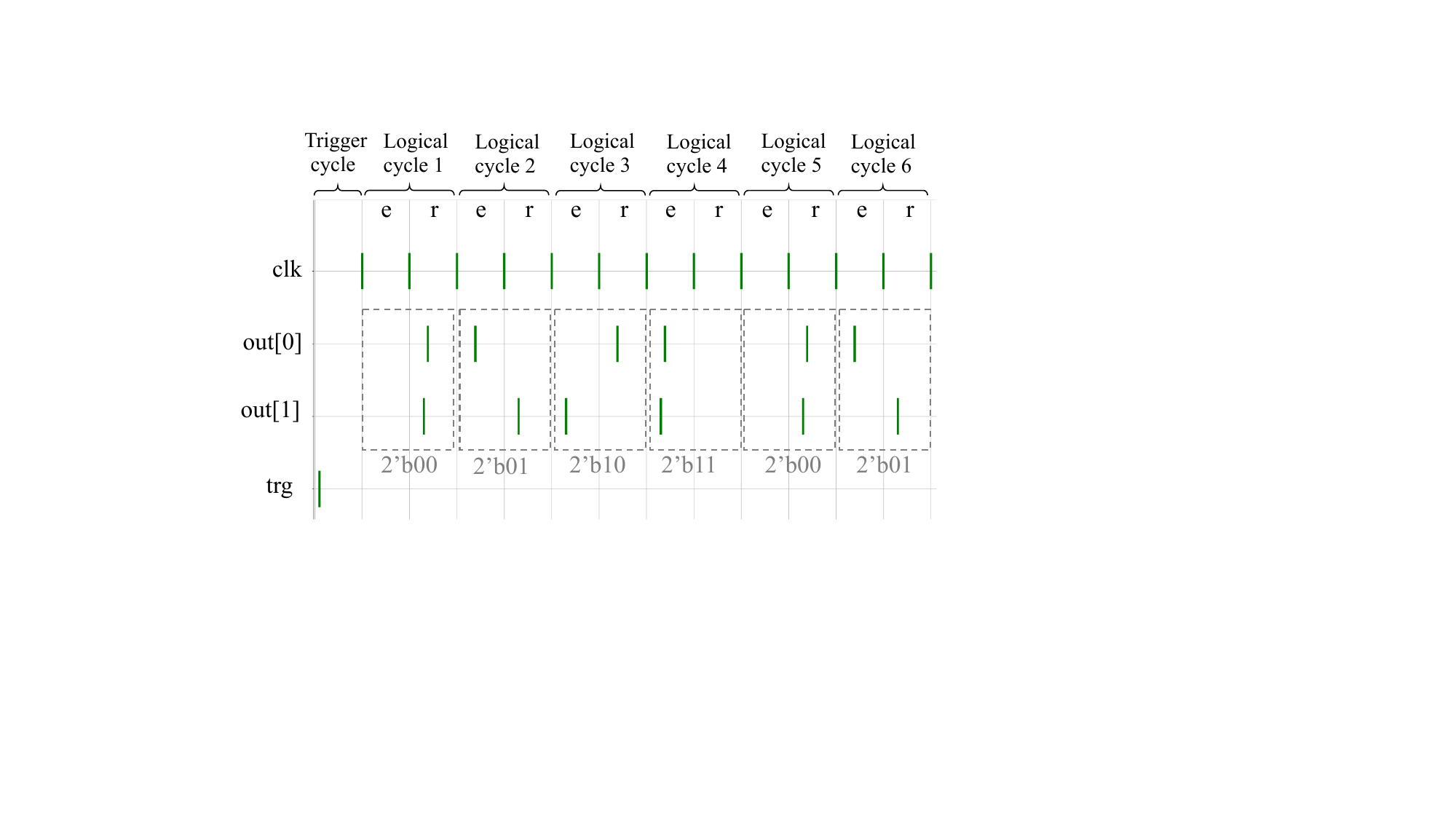}
    \vspace{-2mm}
    \caption{Pulse-level simulation of a 2-bit xSFQ counter in PyLSE~\cite{pylse}. A trigger (trg) signal is provided to set the circuit in the correct state before normal operation starts. A succession of logical cycles, each comprising a pair of synchronous excite (e) and relax (r) phases, then follows. The counter starts at value 00, increments to 11, returns to 00, etc.}
    \label{fig:enter-label}\vspace{-2ex}
\end{figure}

As a solution, we propose to incorporate a separate trigger signal that fires once before the circuit begins operation and propagates to the clock ports of the first (preloaded) DROC cells of every logical flip-flop. This is depicted in Figure~\ref{fig:sync}iii in green. The trigger forces all LA/FA cells in $f_1$ to go into their excite phases, while the LA/FA cells in $f_2$ remain in the init phases. All triggered DROCs reset. The remaining DROCs will receive the outputs of $f_1$. In the next clock cycle, the first rank of DROCs will produce complementary signals and compel $f_1$ to go through a relax phase, while the second rank of DROCs will forward $f_1$'s previous output, as illustrated in Figure~\ref{fig:sync}iv. After this initial cycle, the circuit is ready for operation and all DROCs can be clocked as usual. Simulation results of the proposed methodology for a 2-bit counter are provided in Figure~\ref{fig:enter-label}.

In terms of added implementation costs for the clock tree, only a merger cell (5 JJs) and an external trigger signal, are required. In any case, the xSFQ clock tree will be considerably simpler compared to traditional SFQ approaches. This simplicity arises because just the DROC cells need synchronization and their number is configurable. In other words, the demand for broadcasting a clock signal to all logic gates and to the delay path-balancing DRO cells is eliminated.

%% file: sections/evaluation.tex
\subsection{Setup}
Building upon the results presented in Section~\ref{sec:sfq_lib}, we proceed to apply our synthesis techniques to ISCAS85, EPFL, and ISCAS89 circuits. To this end, we use Yosys v0.33 equipped with ABC v1.01. As described in Section~\ref{sec:synthesis}, no customizations to ABC are needed. We apply only the heuristic described in Section~\ref{sec:domino}.

The above three benchmark suites encompass both combinational and sequential designs. While we have synthesized all the provided circuits in our flow, we present only those that enable comparisons with the state of the art, such as PBMap~\cite{pbmap} and qSeq~\cite{qseq}. Neither PBMap nor qSeq considers PTL interfaces or reports associated costs. For a more equitable comparison, we use the xSFQ figures without PTLs presented in Table~\ref{tab:cells}.

\subsection{Post-synthesis results}
\subsubsection{ISCAS85 \& EPFL combinational circuits}
\label{sec:eval_comb}
A detailed component breakdown and comparison against PBMap~\cite{pbmap} is provided in Table~\ref{tab:comp1}. Notably, unlike RSFQ implementations, our synthesized xSFQ designs feature no DROC or other synchronous cells; therefore, no clock tree is needed. To the best of our understanding, the JJ counts reported in PBMap do not account for this significant overhead. 
Every RSFQ cell requires an additional splitter for its clocking, where every splitter comprises 3 JJs---30\% extra for RSFQ logic cells and 60\% extra for DRO cells used for delay path balancing, even without considering routing costs. We report JJ savings---$4.5\times$ and $5.9\times$ on average---in Table~\ref{tab:comp1}, both without and with a 30\% overhead for clock splitting.

\begin{table}[]
\caption{Post-synthesis component breakdown for ISCAS85 and EPFL circuits and JJ counts compared to PBMap~\cite{pbmap}. JJ savings without/with clock splitting are reported.}
\vspace{-2mm}
\footnotesize
\setlength{\tabcolsep}{3pt} 
\begin{tabular}{|l| S[table-format = 6.0] | S[table-format = 4.0] | S[table-format = 2.0] |c | S[table-format = 5.0] | c|}
\hline
\multirow{2}{*}{\textbf{Circuit}} & \textbf{PBMap} & \multicolumn{5}{c|}{\textbf{This paper}} \\ \cline{2-7} 
& \textbf{\#JJ} & \multicolumn{1}{c|}{\textbf{\#LA/FA}} & \multicolumn{1}{c|}{\textbf{Dupl.}} & \multicolumn{1}{c|}{\textbf{\#DROC}} & \multicolumn{1}{c|}{\textbf{\#JJ}} & \textbf{JJ Savings} \\ \hline
c880 & 12,909 & 452 & 50\% & 0 & 2,942 & 4.4/5.7$\times$ \\ \hline
c1908 & 12,013 & 503 & 71\% & 0 & 3,398 & 3.6/4.6$\times$ \\ \hline
c499 & 7,758 & 682 & 75\% & 0 & 4,624 & 1.7/2.2$\times$ \\ \hline
c3540 & 28,300 & 1,646 & 93\% & 0 & 11,288 & 2.5/3.3$\times$ \\ \hline
c5315 & 52,033 & 1,944 & 42\% & 0 & 13,197 & 4.0/5.1$\times$ \\ \hline
c7752 & 48,482 & 2,571 & 76\% & 0 & 17,157 & 2.8/3.7$\times$ \\ \hline
int2float & 6,432 & 225 & 6\% & 0 & 1,530 & 4.2/5.5$\times$ \\ \hline
dec & 5,469 & 304 & 0\% & 0 & 2,848 & 1.9/2.5$\times$ \\ \hline
priority & 102,085 & 892 & 22\% & 0 & 5,503 & 18.6/24.1$\times$ \\ \hline
sin & 215,318 & 9,977 & 99\% & 0 & 69,770 & 3.1/4.0$\times$ \\ \hline
cavlc & 16,339 & 721 & 8\% & 0 & 5,020 & 3.3/4.2$\times$ \\ \hline
\end{tabular} 
\label{tab:comp1}
\vspace{-3ex}
\end{table}

\subsubsection{ISCAS85 pipelined circuits}

The objective of this section is to present evidence for the effectiveness of xSFQ circuit pipelining. To address pipeline imbalances arising from the use of pairs of DROC cells as logical flip-flops, as discussed in Section~\ref{sec:synthesis}, we utilize ABC's built-in retiming optimizations with default settings. 

Our post-synthesis results for the c6288 circuit from the ISCAS85 benchmark suite are summarized in Table~\ref{tab:pipl}. This circuit was selected from the others due to its size and relatively long logical depth---90 logic gates and 80 splitters on its critical path. The first row indicates no pipelining; the second and third feature one and two pipeline stages, respectively. 

Each pipeline stage introduces two ranks of DROCs due to xSFQ's alternating encoding. Thus, we report both the circuit and architectural clock frequencies for all three cases. The architectural frequency is half that of the circuit---both excite and relax phases must be processed for every input. For higher speeds, more pipeline stages should be added. Notably, the number of JJs increases sub-linearly with the number of DROCs for pipelining. This happens because the addition of pipeline stages creates more opportunities to apply the optimizations discussed in Section~\ref{sec:synthesis}.

\begin{table}[]
\caption{Post-synthesis results for c6288 from the ISCAS85 benchmark suite. Two numbers are reported for the number of pipeline stages (architectural/circuit), DROC cells (without/with preloading), logical depth (without/with splitters), and clock frequency (circuit/architectural).}
\vspace{-2mm}
\footnotesize
\setlength{\tabcolsep}{3pt} 
\begin{tabular}{|c|c|c|c|c|c|c|}
\hline
\textbf{\begin{tabular}[c]{@{}c@{}}\# Pipeline \\ stages\end{tabular}} & \textbf{\#JJ} & \textbf{\#LA/FA} & \textbf{Dupl.} & \textbf{\#DROC} & \textbf{\begin{tabular}[c]{@{}c@{}}Logical \\ depth\end{tabular}} & \textbf{\begin{tabular}[c]{@{}c@{}}Clock \\ freq. (GHz)\end{tabular}} \\ \hline
$0/0$                                                                  & 25,853        & 3,707            & 97\%           & 0/0             & $90/170$                                                          & $0.9/0.5$                                                             \\ \hline
$1/2$                                                                  & 27,312        & 3,669            & 95\%           & 91/32           & $46/90$                                                           & $1.6/0.8$                                                             \\ \hline
$2/4$                                                                  & 29,399        & 3,572            & 89\%           & 171/123         & $24/48$                                                           & $3.0/1.5$                                                             \\ \hline
\end{tabular}
\label{tab:pipl}
\vspace{-2ex}
\end{table}

\vspace{-1mm}
\begin{table}[t]
\caption{Post-synthesis component breakdown for ISCAS89 circuits and JJ counts compared to qSeq~\cite{qseq}.}
\vspace{-2mm}
\footnotesize
\setlength{\tabcolsep}{3pt} 
\begin{tabular}{| l | S[table-format = 5.0] | S[table-format = 3.0] | c | c | S[table-format = 4.0] | c|}
\hline
\multirow{2}{*}{\textbf{Circuit}} & \textbf{qSeq}  & \multicolumn{5}{c|}{\textbf{This paper}}                                                                                                                                       \\ \cline{2-7} 
                                  & \textbf{\#JJs} & \multicolumn{1}{c|}{\textbf{\#LAs/FAs}} & \multicolumn{1}{c|}{\textbf{Dupl.}} & \multicolumn{1}{c|}{\textbf{\#DROCs}} & \multicolumn{1}{c|}{\textbf{\#JJs}} & \textbf{JJ Savings} \\ \hline
s27 & 527 & 12 & 71\% & 3/3 & 162 & 3.3/4.3$\times$ \\ \hline
s298 & 3,698 & 107 & 24\% & 18/14 & 1,228 & 3.0/3.9$\times$ \\ \hline
s344 & 5,475 & 117 & 24\% & 19/15 & 1,357 & 4.0/5.2$\times$ \\ \hline
s349 & 5,475 & 118 & 26\% & 19/15 & 1,364 & 4.0/5.2$\times$ \\ \hline
s382 & 4,934 & 135 & 26\% & 29/21 & 1,724 & 2.9/3.8$\times$ \\ \hline
s386 & 4,580 & 153 & 61\% & 11/6 & 1,295 & 3.5/4.6$\times$ \\ \hline
s400 & 5,144 & 133 & 30\% & 25/21 & 1,664 & 3.1/4.0$\times$ \\ \hline
s420.1 & 5,661 & 128 & 20\% & 16/16 & 1,354 & 4.2/5.5$\times$ \\ \hline
s444 & 5,148 & 133 & 36\% & 28/21 & 1,706 & 3.0/3.9$\times$ \\ \hline
s510 & 7,085 & 287 & 31\% & 19/6 & 2,265 & 3.1/4.0$\times$ \\ \hline
s526 & 6,365 & 159 & 24\% & 25/21 & 1,819 & 3.5/4.6$\times$ \\ \hline
s641 & 11,462 & 167 & 34\% & 17/17 & 1,653 & 6.9/9.0$\times$ \\ \hline
s713 & 11,421 & 167 & 34\% & 17/17 & 1,653 & 6.9/9.0$\times$ \\ \hline
s820 & 9,797 & 308 & 34\% & 6/5 & 2,284 & 4.3/5.6$\times$ \\ \hline
s832 & 9,641 & 298 & 32\% & 5/5 & 2,204 & 4.4/5.7$\times$ \\ \hline
s838.1 & 12,710 & 256 & 17\% & 32/32 & 2,714 & 4.7/6.1$\times$ \\ \hline
\end{tabular}
\label{tab:seq}
\vspace{-4ex}
\end{table}

\subsubsection{ISCAS89 sequential circuits} Similar to Section~\ref{sec:eval_comb}, we present a post-synthesis component breakdown and comparisons against the state of the art. The difference is that here we examine ISCAS89 circuits, which are sequential. Thus, we use qSeq~\cite{qseq} as a comparison point instead of PBMap~\cite{pbmap}, which again, to the best of our knowledge, does not consider clock tree or routing costs. The results are summarized in Table~\ref{tab:seq}. On average, our circuits require $4.1\times$ and $5.3\times$ fewer JJs than qSeq, without and with a 30\% overhead for clock splitting.

%% file: sections/conclusion.tex
This paper introduces a comprehensive methodology for synthesizing resource-efficient superconducting circuits using established techniques and off-the-shelf tools. To do so, we leverage xSFQ, a recently introduced logic family that removes the clock from superconducting gate semantics. We firstly redesign xSFQ's logic and storage cells to enhance cascadability and create the conditions necessary to exploit a new link with domino logic. Then, we build an xSFQ cell library by characterizing the above designs in Synopsys' HSPICE and utilize unmodified Yosys and ABC tools to optimize both combinational and sequential circuits from ISCAS85, EPFL, and ISCAS89 benchmark suites. Our post-synthesis results indicate a reduction in the number of required JJs ranging from 2 to 24 times compared to the state of the art, even under conservative assumptions. In a broader context, these accomplishments denote notable progress, not only in enhancing logic density in superconducting circuits but also in the simplification of addressing design automation challenges in non-latching emerging technologies. This simplification is rooted in the targeted exploitation of new logic advancements, which contrasts prior methods that rely on brute force or ad hoc solutions.